\documentclass[aps,floats]{revtex4}
\usepackage{amsmath,amssymb}
\usepackage{graphicx,epsfig}

\begin{document}
\bibliographystyle {plain}

\def\oppropto{\mathop{\propto}} 
\def\opsimeq{\mathop{\simeq}}
\def\opoverderline{\mathop{\overline}}
\def\operarrow{\mathop{\longrightarrow}}
\def\opsim{\mathop{\sim}}

\def\fig#1#2{\includegraphics[height=#1]{#2}}
\def\figx#1#2{\includegraphics[width=#1]{#2}}


\title{ Chaos properties of the one-dimensional long-range Ising spin-glass } 


 \author{ C\'ecile Monthus and Thomas Garel }
  \affiliation{ Institut de Physique Th\'{e}orique, CNRS and CEA Saclay,
 91191 Gif-sur-Yvette, France}

\begin{abstract}

For the long-range one-dimensional Ising spin-glass with random couplings decaying as $J(r) \propto r^{-\sigma}$, the scaling of the effective coupling defined as the difference between the free-energies corresponding to Periodic and Antiperiodic boundary conditions $J^R(N) \equiv  F^{(P)}(N)-F^{(AP)}(N) \sim N^{\theta(\sigma)}$ defines the droplet exponent $\theta(\sigma)$. Here we study numerically the instability of the renormalization flow of the effective coupling $J^R(N)$ with respect to magnetic, disorder and temperature perturbations respectively, in order to extract the corresponding chaos exponents $\zeta_H(\sigma)$, $\zeta_J(\sigma)$ and $\zeta_T(\sigma)$ as a function of $\sigma$. Our results for $\zeta_T(\sigma) $ are interpreted in terms of the entropy exponent $\theta_S(\sigma) \simeq 1/3$ which governs the scaling of the entropy difference $ S^{(P)}(N)-S^{(AP)}(N) \sim N^{\theta_S(\sigma)}$. We also study the instability of the ground state configuration with respect to perturbations, as measured by the spin overlap between the unperturbed and the perturbed ground states, in order to extract the corresponding chaos exponents $\zeta^{overlap}_H(\sigma)$ and $\zeta^{overlap}_J(\sigma)$. 

\end{abstract}

\maketitle

\section{ Introduction }

\subsection{ Chaos as instability of the renormalization flow}

In the field of dynamical systems, the notion of chaos means `sensitivity
to initial conditions' and is quantified by the Lyapunov exponent $\lambda >0$
which governs the exponential growth of the distance between two dynamical trajectories
$\delta(t) \propto e^{\lambda t} \delta(0)$ that are separated by an infinitesimal
distance $\delta(0)$  at time $t=0$.
In the field of spin-glasses, the notion of 'chaos' has been introduced as the sensitivity of the renormalization flow seen as a 'dynamical system', with respect to the initial conditions
(the random couplings) or with respect to external parameters like the temperature $T$
or the magnetic field $H$. On hierarchical lattices where explicit renormalization rules exist for the renormalized couplings $J^R$, the chaos properties have been thus much studied \cite{berker,brayM,brayB,cieplak,hilhorst,aspelmeierBM,sasakiMK,krzakala,jorg,bresil}.
For other lattices without explicit renormalization rules, the droplet scaling theory
\cite{mcmillan,bray_moore,fisher_huse} allows to define the chaos properties as follows :

(i) the effective renormalized coupling $J^R$ of a $d$-dimensional disordered sample of linear size $L$ containing $N=L^d$ spins
can be defined as the difference between the free-energies $F^{(P)}(N)$ and $F^{(AP)}(N)$
corresponding to Periodic and Antiperiodic boundary conditions in the first direction respectively (the other $(d-1)$ directions keep periodic boundary conditions)
\begin{eqnarray}
J^R(N=L^d) \equiv  F^{(P)}(N)-F^{(AP)}(N)  = L^{\theta^{linear}} u = N^{\theta} u
\label{jrtheta}
\end{eqnarray}
where $\theta^{linear}$ is the usual 
droplet exponent associated to the linear size $L$
, and where $u$ is an $O(1)$ random variable of zero mean (with a probability distribution symmetric in $u \to -u$). In the following, we will use the droplet exponent $\theta=\theta^{linear}/d$ defined here with respect to  the total number $N$ of spins, in order to consider also fully connected models where the notion of length does not exist. 

(ii) for the same disordered sample, 
one may now consider a perturbation $\delta$ 
(either in the disorder, temperature or magnetic field) and
the corresponding renormalized coupling
\begin{eqnarray}
J^R_{\delta}(N) \equiv  F^{(P)}_{\delta}(N)-F^{(AP)}_{\delta} (N) 
\label{jrthetaper}
\end{eqnarray}
and construct the disorder-averaged correlation function
\begin{eqnarray}
C_{\delta}(N) \equiv \frac{ \overline{J^R(N) J^R_{\delta}(N)}}
{\sqrt{\overline{(J^R(N))^2}}
\sqrt{\overline{(J^R_{\delta}(N))^2}}}
\label{corre}
\end{eqnarray}
The chaos exponent $\zeta_{\delta}$ associated to the perturbation $\delta$
is then defined by the size dependence of the decorrelation scale at small perturbation $\delta$
\begin{eqnarray}
C_{\delta}(N) \opsimeq_{\delta \to 0} 1- a (\delta N^{\zeta_{\delta}})^2+o(\delta^2)
\label{corredefzeta}
\end{eqnarray}
where $a$ is a numerical constant.
This method has been used to measure numerically the chaos exponents for spin-glasses on hypercubic lattices \cite{brayM,aspelmeierBM,sasaki,lukic,middleton}. 
For each type of perturbation $\delta$ (magnetic, disorder, temperature), the droplet scaling theory predicts values of the corresponding chaos exponent $\zeta_{\delta}$
\cite{fisher_huse,brayM}, as will be recalled below in the text.

\subsection{ Chaos as instability of the spin configurations } 

Besides the scaling droplet theory of spin-glasses recalled above,
the alternative Replica-Symmetry-Breaking scenario \cite{replica}
based on the mean-field fully connected Sherrington-Kirkpatrick model \cite{SKmodel}
considers that the main observable of the spin-glass phase is the overlap between configurations. As a consequence, another notion of chaos as been introduced 
\cite{parisi,kondor,ritort,ritort2} based on the overlap between the spins $S_i^{(0)}$ of the unperturbed system 
and the spins $S_i^{(\delta)}$ of the perturbed system 
\begin{eqnarray}
 q_{(0,\delta)}(N) \equiv \frac{1}{N} \left\vert \sum_{i=1}^N S_i^{(0)} S_i^{(\delta)} \right\vert
\label{qdelta}
\end{eqnarray}
The dimensionless 'chaoticity parameter' \cite{ritort}
\begin{eqnarray}
 r_{\delta}(N) \equiv \frac{ \overline{ <q_{(0,\delta)}(N)> } }
{ \sqrt{\overline{ <q_{(0,0)}(N)>}} \sqrt{\overline{ <q_{(\delta,\delta)}(N)>}}}
\label{chaoticity}
\end{eqnarray}
has been much studied  in various spin-glass models
\cite{ritort,ritort2,franz,rieger,muriel,BilloireChaos,barbara,rizzo,krzakalaJPB,katz_krzakala,aspelmeier,parisi_rizzo,fernandez}
in order to extract the chaos exponent $\zeta_{\delta}^{overlap}$ that governs
the size dependence of the decorrelation scale at small perturbation $\delta$
\begin{eqnarray}
r_{\delta}(N) \opsimeq_{\delta \to 0} 1- b \delta N^{\zeta^{overlap}_{\delta}}+o(\delta)
\label{corredefzetaover}
\end{eqnarray}
where $b$ is a numerical constant.

\subsection{ Organization of the paper }

The aim of this work is to study the chaos properties of the one-dimensional long-range Ising spin-glass with respect to various perturbations, using the two procedures described above. 
The paper is organized as follows.
In section \ref{sec_LRSG}, we recall the properties of the one-dimensional long-range Ising spin-glass.
The chaos exponents based on the correlation of Eq. \ref{corre}
are studied for magnetic, disorder and temperature perturbations
in sections \ref{sec_magnetic}, \ref{sec_disorder} and \ref{sec_temp}
respectively. The instability of the ground-state with respect to magnetic and disorder perturbations as measured by the chaoticity parameter of Eq. \ref{chaoticity}
is analyzed in section \ref{sec_overlap}.
Our conclusions are summarized in section \ref{sec_conclusion}.
In Appendix \ref{app_hlocmin}, we discuss the scaling of the lowest local field
as a function of the system size, in order to interpret the results 
found in section \ref{sec_overlap}.

\section{ Reminder on the one-dimensional long-range Ising spin-glass  }

\label{sec_LRSG}

The one-dimensional long-range Ising spin-glass introduced in \cite{kotliar}
  allows to interpolate continuously between the one-dimensional nearest-neighbor model and the Sherrington-Kirkpatrick mean-field model \cite{SKmodel}. Since it is much simpler to study numerically than hypercubic lattices as a function of the dimension $d$, 
 this model has attracted a lot of interest recently \cite{KY,KYgeom,KKLH,KKLJH,Katz,KYalmeida,Yalmeida,KDYalmeida,LRmoore,KHY,KH,mori,wittmann,us_overlaptyp,us_dynamic}
(here we will not consider the diluted version of the model \cite{diluted}).

\subsection {Definition of the model}

The one-dimensional long-range Ising spin-glass \cite{kotliar} is defined by the Hamiltonian
\begin{eqnarray}
 {\cal H} && = - \sum_{1 \leq i <j \leq N} J_{ij} S_i S_j
\label{HSGring}
\end{eqnarray}
where the $N$ spins $S_i=\pm 1$ lie periodically on a ring, so that the distance $r_{ij} $ between the spins $S_i$ and $S_j$ reads \cite{KY}
\begin{eqnarray}
r_{ij}= \frac{N}{\pi} \sin \left(\vert j-i \vert \frac{\pi}{N} \right)
\label{rij}
\end{eqnarray}
The couplings are chosen to decay with respect to this distance 
as a power-law of exponent $\sigma$
\begin{eqnarray}
J_{ij}= c_N(\sigma) \frac{\epsilon_{ij}}{r_{ij}^{\sigma}}
\label{defjij}
\end{eqnarray}
where $\epsilon_{ij}$ are random Gaussian variables
of zero mean $\overline{\epsilon}=0$ and unit variance $\overline{\epsilon^2}=1 $.
The constant $c_N(\sigma) $ is defined by the condition \cite{KY}
\begin{eqnarray}
1= \sum_{j \ne 1} \overline{J_{1j}^2} =  c_N^2(\sigma) \sum_{j \ne 1} \frac{1}{r_{1j}^{2 \sigma}}
\label{defcsigma}
\end{eqnarray}
that ensures the extensivity of the energy.
The exponent $\sigma$ is thus the important parameter of the model.

\subsection {Periodic versus Antiperiodic boundary conditions }

\label{subsec_anti}

For the long-range model of Eq. \ref{HSGring}, 
'Antiperiodic boundary conditions' means the following prescription \cite{KY} :
for each disordered sample $(J_{ij})$ considered as 'Periodic', the 'Antiperiodic'
consists in changing the sign $J_{ij} \to -J_{ij}$ for all pairs $(i,j)$ where
the shortest path on the circle goes through the bond $(L,1)$.

\subsection{Non-extensive region  $0 \leq \sigma < 1/2$ } 

\label{nonextregion}

In the non-extensive region $0 \leq \sigma < 1/2$, 
Eq. \ref{defcsigma} yields 
\begin{eqnarray}
 c_N(\sigma) \propto N^{\sigma- \frac{1}{2}}
\label{rescalcsigma}
\end{eqnarray}
so there is an explicit size-rescaling of the couplings
as in the Sherrington-Kirkpatrick (SK)
mean-field model \cite{SKmodel} which corresponds to the case $\sigma=0$.
Recent studies \cite{mori,wittmann} have proposed that
both universal properties like critical exponents, but also
non-universal properties like the critical temperature 
do not depend on $\sigma$ in the whole region $0 \leq \sigma < 1/2$,
and thus coincide with the properties of the SK model $\sigma=0$.
For the SK model $\sigma=0$, there seems to be a consensus on the shift
 exponent governing the correction to extensivity of
the averaged value ground state energy \cite{andreanov,Bou_Krz_Mar,pala_gs,aspelmeier_MY,Katz_gs,Katz_guiding,aspelmeier_BMM,boettcher_gs,us_tails,us_matching}
\begin{eqnarray}
 \theta_{shift} (\sigma) \simeq 1/3
\label{shiftSK}
\end{eqnarray}
The droplet exponent $\theta(\sigma)$ measured via Eq \ref{jrtheta}
in Ref \cite{KY}
is indeed compatible with this constant value in the whole non-extensive region
\begin{eqnarray}
\theta(0 \leq \sigma < 1/2)  \simeq 1/3
\label{thetanonext}
\end{eqnarray}

\subsection{Extensive region $\sigma>1/2$  } 

\label{extregion}

In the extensive region $\sigma>1/2$, Eq. \ref{defcsigma} yields 
\begin{eqnarray}
 c_N(\sigma) =O(1)
\label{norescalcsigma}
\end{eqnarray}
so that there is no size rescaling of the couplings.
The limit $\sigma =+\infty$ corresponds to the nearest-neighbor one-dimensional model.  The droplet exponent $\theta(\sigma)$ has been measured using Eq \ref{jrtheta}
 via Monte-Carlo simulations on sizes $L \leq 256$
with the following results \cite{KY}
(see \cite{KY} for other values of $\sigma$)
\begin{eqnarray}
\theta(\sigma=0.62) && \simeq 0.24
\nonumber \\
\theta(\sigma=0.75) && \simeq 0.17
\nonumber \\
\theta(\sigma=0.87) && \simeq 0.08
\nonumber \\
\theta(\sigma=1) && \simeq 0
\nonumber \\
\theta(\sigma=1.25) && \simeq -0.24
\label{thetadw1d}
\end{eqnarray}
In our previous work \cite{us_overlaptyp}, we have found
that exact enumeration on much smaller sizes $6 \leq L \leq 24$
 actually yield values close to Eq. \ref{thetadw1d}.

There exists a spin-glass phase at low temperature for $\sigma<1$ \cite{kotliar}, characterized by a positive droplet exponent $\theta(\sigma)>0$.

\section{ Magnetic field chaos exponent $\zeta_H(\sigma)$ }

\label{sec_magnetic}

In the presence of an external magnetic field $H$, the Hamiltonian of Eq. \ref{HSGring} becomes
\begin{eqnarray}
 {\cal H}_H && = - \sum_{1 \leq i <j \leq N} J_{ij} S_i S_j-H \sum_{i=1}^{N} S_i
\label{HSGringmagn}
\end{eqnarray}

\subsection{ Scaling prediction of the droplet theory}

Within the droplet scaling theory \cite{bray_moore,fisher_huse},
the chaos exponent associated to a magnetic field perturbation $H$
can be predicted via the following Imry-Ma argument : 
 the field $H$ couples to the random magnetization of order $ N^{1/2}$ of the 
extensive droplet of the unperturbed spin-glass state. The induced perturbation 
of order
\begin{eqnarray}
 \Delta_H(N) \propto H N^{1/2}
\label{Hper}
\end{eqnarray}
has to be compared with the renormalized coupling $J^R(N) \sim N^{\theta} u$
 of Eq. \ref{jrtheta}. The appropriate scaling parameter is thus 
$H N^{\zeta_H}$ with the magnetic field chaos exponent
\begin{eqnarray}
 \zeta_H=\frac{1}{2}-\theta
\label{zetaH}
\end{eqnarray}

\subsection{ Numerical results for the long-range Ising spin-glass }

\begin{figure}[htbp]
\includegraphics[height=6cm]{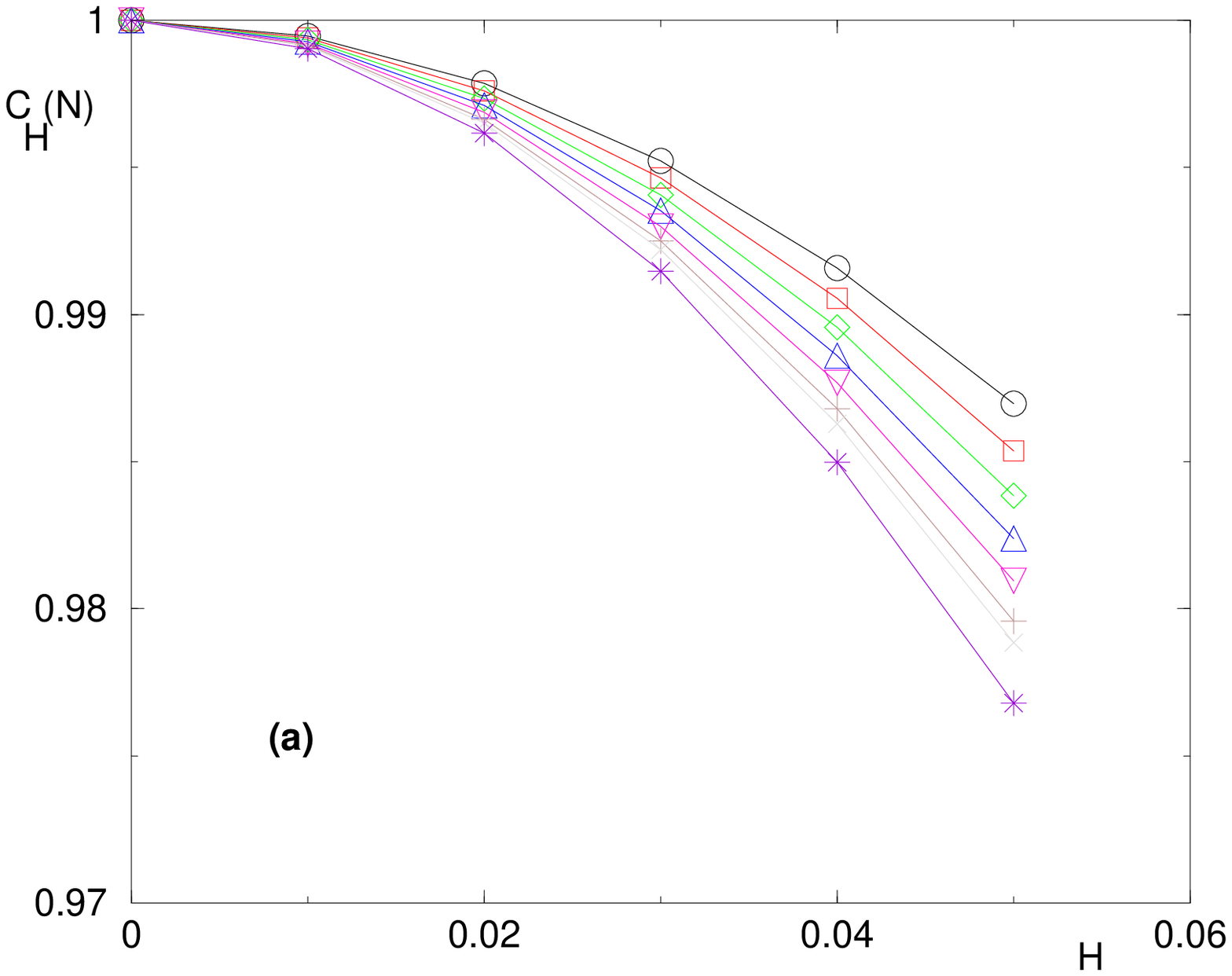}
\hspace{1cm}
 \includegraphics[height=6cm]{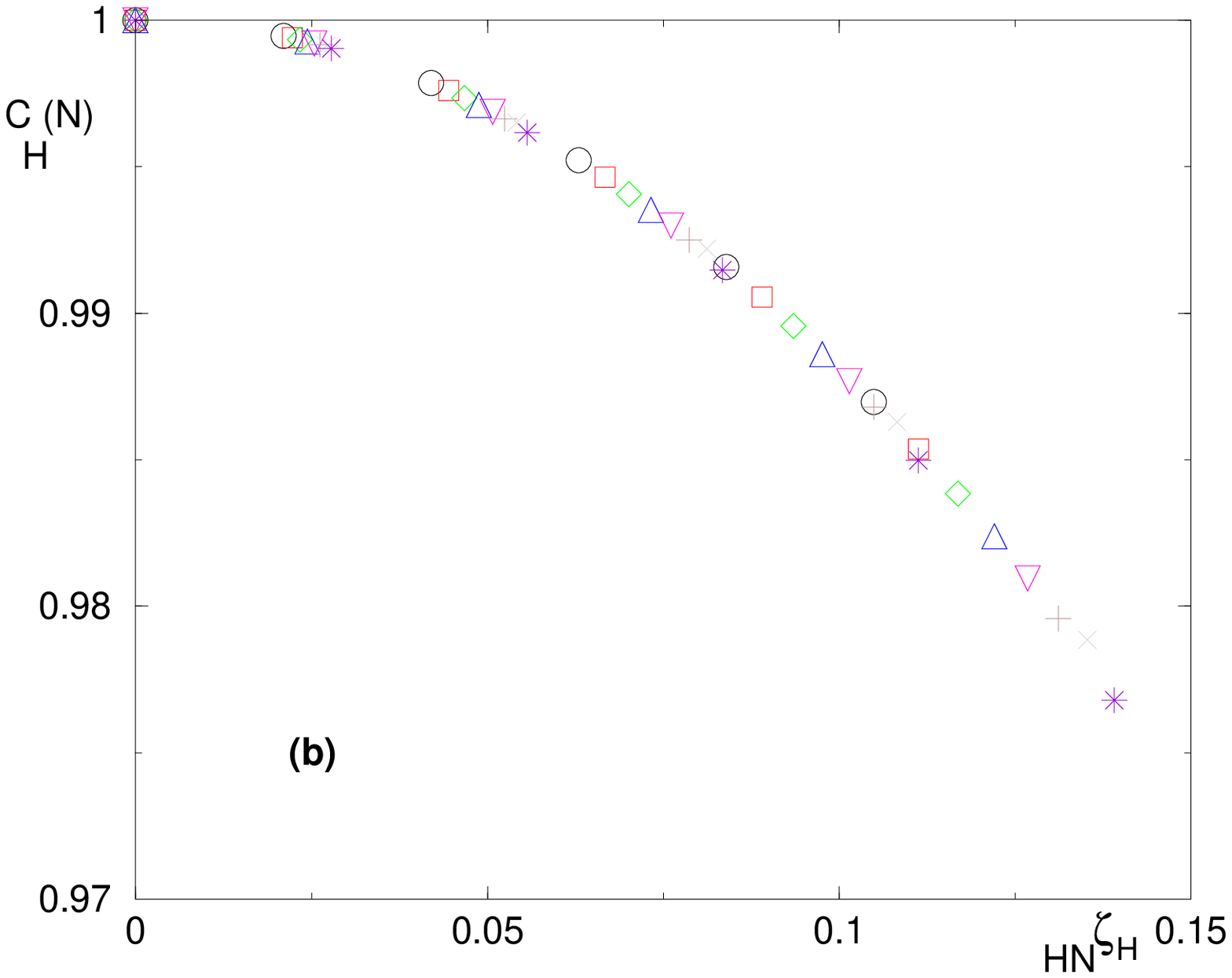}
\caption{ Measure of the magnetic chaos exponent $\zeta_H$ for $\sigma=0.75$ :
(a) Results for the correlation $C_{H}(N)$ as a function of the magnetic field 
$H=0.01,0.02,0.03,0.04,0.05$ for various sizes $10 \leq N \leq 24$.
(b) Same data as a function of the rescaled variable $H N^{\zeta_H}$
with $\zeta_H(\sigma=0.75)  \simeq 0.32 $. }
\label{figmagne}
\end{figure}

We have measured the correlation $C_{H}(N)$ of Eq. \ref{corre} 
at zero temperature $T=0$, so that the free-energies $F$ in Eq. \ref{jrthetaper}
corresponds to the ground state energy $E^{GS}$
\begin{eqnarray}
C_{H}(N) \equiv \frac{ \overline{\left[ E^{GS(P)}_{H=0}(N)-E^{GS(AP)}_{H=0} (N)\right]\left[ E^{GS(P)}_{H}(N)-E^{GS(AP)}_{H} (N)\right] }}
{\sqrt{\overline{\left[ E^{GS(P)}_{H=0}(N)-E^{GS(AP)}_{H=0} (N)\right]^2}}
\sqrt{\overline{\left[ E^{GS(P)}_{H}(N)-E^{GS(AP)}_{H} (N)\right]^2}}}
\label{correHzeroT}
\end{eqnarray}
The ground state energies $E^{GS(P)}_{H}(N) $ and $ E^{GS(AP)}_{H} (N)$ 
corresponding to Periodic or Antiperiodic boundary conditions
for various values of the external magnetic field $H$ have been measured
via exact enumeration
of the $2^N$ spin configurations for small even sizes $10 \leq N \leq 24$.
 The statistics over samples have been obtained
for instance with the following numbers $n_s(N)$ of disordered samples
\begin{eqnarray}
n_s(L \leq 10) =10^9 ; ...;n_s(L = 16)=53.10^5 ; ... ;  n_s(L = 24)=12\times10^3 
\label{ns1copy}
\end{eqnarray}
We have used five small values of the magnetic field
$H=0.01,0.02,0.03,0.04,0.05$ in order to extract the chaos exponent from
the expansion (Eq \ref{corredefzeta})
\begin{eqnarray}
C_{H}(N) \opsimeq_{\delta \to 0} 1- a_{magnetic} (H N^{\zeta_{H}})^2+o(H^2)
\label{corredefzetaH}
\end{eqnarray}
where $a_{magnetic} $ is a numerical constant.
As an example, we show on Fig. \ref{figmagne} our data for $\sigma=0.75$.

In the non-extensive region $0 \leq \sigma <1/2$, our numerical results are compatible
with the value given by Eqs \ref{thetanonext} and \ref{zetaH}
\begin{eqnarray}
 \zeta_H(0 \leq \sigma <1/2)=\frac{1}{2}-\theta(0 \leq \sigma <1/2) \simeq \frac{1}{6} \simeq 0.17
\label{zetaHnonext}
\end{eqnarray}

In the extensive region, our numerical measures as a function of $\sigma$ 
\begin{eqnarray}
\zeta_H(\sigma=0.62) && \simeq 0.26
\nonumber \\
\zeta_H(\sigma=0.75) && \simeq 0.32
\nonumber \\
\zeta_H(\sigma=0.87) && \simeq 0.39
\nonumber \\
\zeta_H(\sigma=1) && \simeq 0.47
\nonumber \\
\zeta_H(\sigma=1.25) && \simeq 0.64
\label{zetaHnum}
\end{eqnarray}
are in reasonable agreement with the formula of Eq. \ref{zetaH} and
the values of the droplet exponent $\theta(\sigma)$ recalled in Eq \ref{thetadw1d}.

\section{Disorder chaos exponent $\zeta_J(\sigma)$ }

\label{sec_disorder}

For each realization of the couplings of Eq. \ref{defjij},
 we draw independent Gaussian random variables $\epsilon_{ij}'$
of zero mean and unit variance, and we consider the following perturbation
of amplitude $\delta$ of the couplings of Eq. \ref{defjij}
\begin{eqnarray}
J^{(\delta)}_{ij}= c_N(\sigma) \frac{\left(\frac{\epsilon_{ij}+\delta \epsilon_{ij}' }{\sqrt{1+\delta^2}} \right)}{r_{ij}^{\sigma}}
\label{defjijnewdelta}
\end{eqnarray}

\subsection{ Scaling prediction of the droplet theory}

Within the droplet scaling theory \cite{bray_moore,fisher_huse},
the chaos exponent associated to a disorder perturbation 
for {\it short-range } spin-glasses
can be predicted via the following Imry-Ma argument : 
 the disorder perturbation of amplitude $\delta$ which couples to the 
{\it surface of dimension $d_s$} of the extensive droplet
\begin{eqnarray}
 \Delta^{SR}_J(N) 
\propto \delta L^{\frac{d_s}{2}} = \delta N^{\frac{d_s}{2d}}
\label{JperSR}
\end{eqnarray}
has to be compared with the renormalized coupling $J^R(N) \sim N^{\theta} u$
 of Eq. \ref{jrtheta}. The appropriate scaling parameter is thus 
$\delta N^{\zeta_J}$ with the disorder chaos exponent
\begin{eqnarray}
 \zeta^{SR}_J=\frac{d_s}{2 d}-\theta
\label{zetaJSR}
\end{eqnarray}

For the long-range one-dimensional model, the scaling of the induced perturbation
has to be re-evaluated from the following double sum involving
 one point $i$ in the droplet $D$ and one point $j$
outside the droplet $D$
\begin{eqnarray}
 \Delta^{LR}_J(N)\propto  \delta c_N(\sigma) 
\sqrt{ \sum_{i \in D} \sum_{j \notin D}   \frac{1}{ \vert i -j \vert^{2 \sigma}} }
\label{JperLR}
\end{eqnarray}

In the non-extensive regime $\sigma<1/2$, the sum is dominated by the
large distances $\vert i -j \vert$, so that taking into account Eq. \ref{rescalcsigma}, Eq. \ref{JperLR} behaves as
\begin{eqnarray}
 \Delta^{(0 \leq \sigma<1/2) }_J(N)\propto  \delta c_N(\sigma) 
\sqrt{ N^{2-2 \sigma } }= \delta N^{\frac{1}{2}}
\label{JperLRnonext}
\end{eqnarray}
yielding the chaos exponent
\begin{eqnarray}
 \zeta^{(0 \leq \sigma<1/2)}_J=\frac{1}{2 }-\theta(\sigma)
\label{zetaJnonext}
\end{eqnarray}

In the extensive regime $\sigma>1/2$, the sum of Eq. \ref{JperLR} is dominated
by the short distances in $\vert i -j \vert$, so that one recovers the 
scaling of Eq. \ref{zetaJSR}
\begin{eqnarray}
 \zeta^{\sigma>1/2}_J=\frac{d_s(\sigma)}{2 }-\theta(\sigma)
\label{zetaJext}
\end{eqnarray}
where the surface dimension $d_s(\sigma)$ is expected to vary
between $d_s(\sigma=1/2)=1$ to match the non-extensive regime of Eq. \ref{zetaJnonext}, and  $d_s(\sigma \to +\infty)=0$ to match the exact result of the one-dimensional nearest-neighbor model \cite{brayM}.

\subsection{ Numerical results for the long-range Ising spin-glass } 

\begin{figure}[htbp]
\includegraphics[height=6cm]{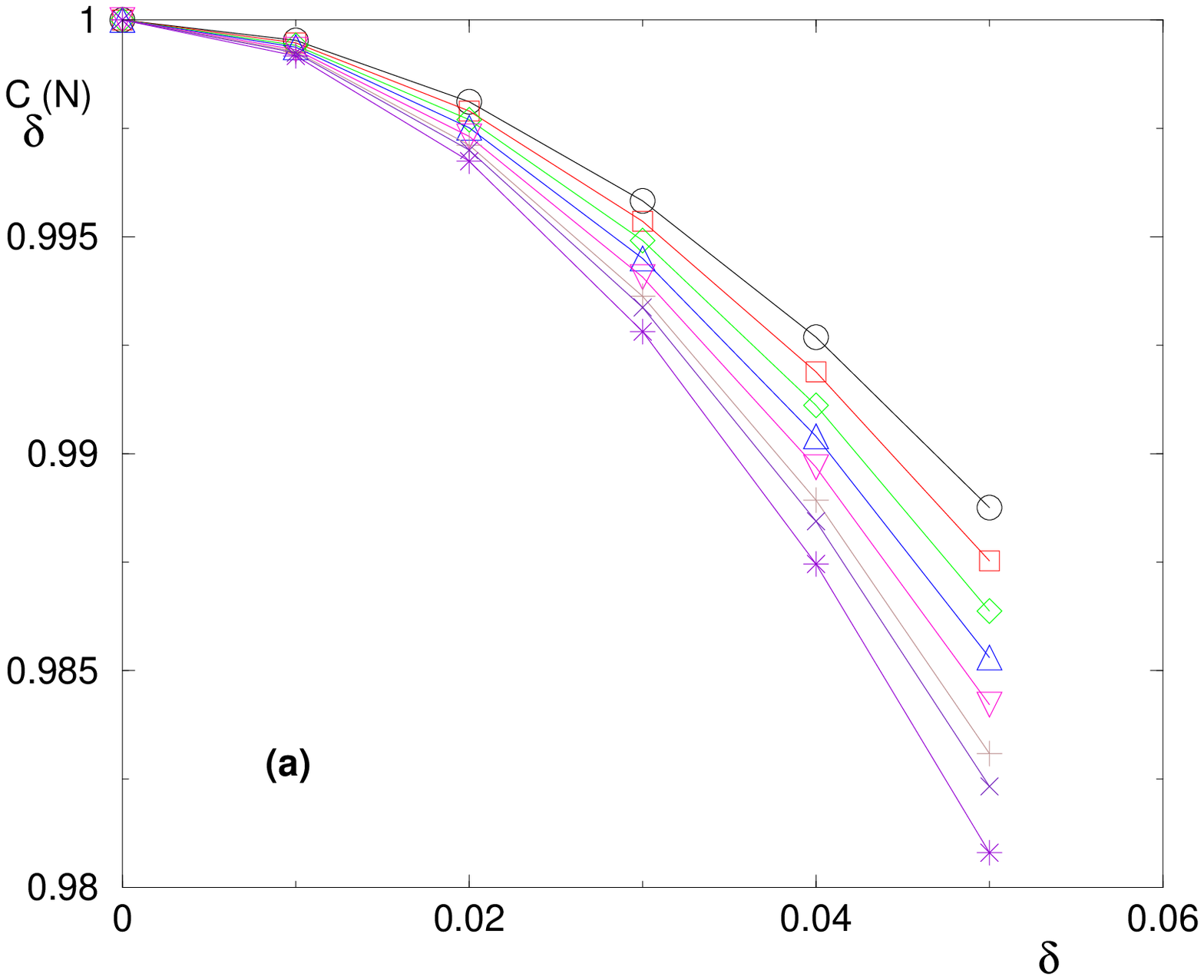}
\hspace{1cm}
 \includegraphics[height=6cm]{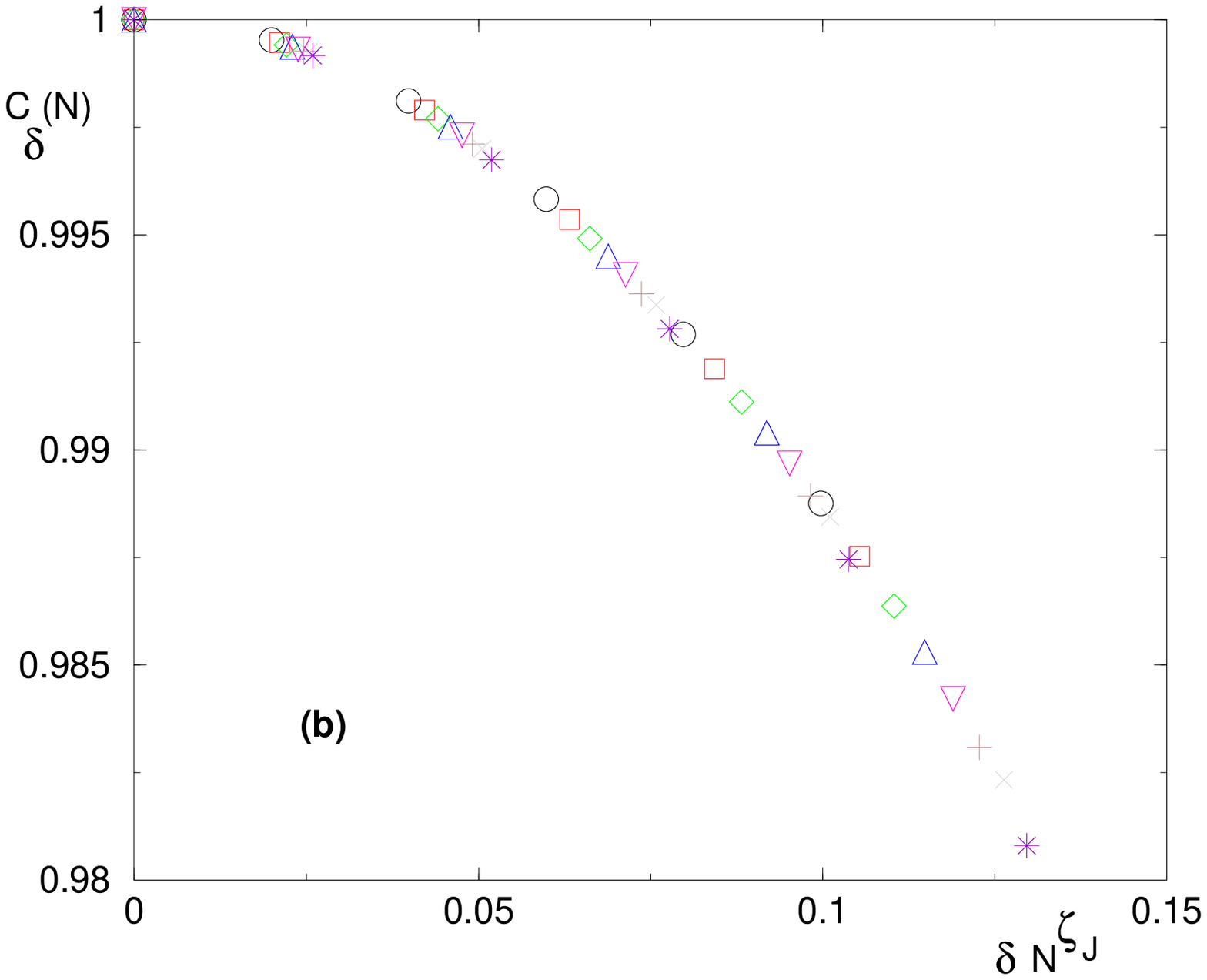}
\caption{ Measure of the disorder chaos exponent $\zeta_J$ for $\sigma=0.75$ :
(a) Results for the correlation $C_{\delta}(N)$ as a function of the amplitude
$\delta=0.01,0.02,0.03,0.04,0.05$ of the perturbation (Eq. \ref{defjijnewdelta})
for various sizes $10 \leq N \leq 24$.
(b) Same data as a function of the rescaled variable $\delta N^{\zeta_J}$
with $\zeta_J(\sigma=0.75)  \simeq 0.3 $. }
\label{figdisorder}
\end{figure}

We have measured the correlation $C_{\delta}(N)$ of Eq. \ref{corre} 
at zero temperature $T=0$, so that the free-energies $F$ in Eq. \ref{jrthetaper}
corresponds to the ground state energy $E^{GS}$
\begin{eqnarray}
C_{\delta}(N) \equiv \frac{ \overline{\left[ E^{GS(P)}_{\delta=0}(N)-E^{GS(AP)}_{\delta=0} (N)\right]\left[ E^{GS(P)}_{\delta}(N)-E^{GS(AP)}_{\delta} (N)\right] }}
{\sqrt{\overline{\left[ E^{GS(P)}_{\delta=0}(N)-E^{GS(AP)}_{\delta=0} (N)\right]^2}}
\sqrt{\overline{\left[ E^{GS(P)}_{\delta}(N)-E^{GS(AP)}_{\delta} (N)\right]^2}}}
\label{corredeltazeroT}
\end{eqnarray}
The ground state energies corresponding to Periodic or Antiperiodic boundary conditions
for various values of the perturbation amplitude $\delta$ of Eq. \ref{defjijnewdelta}
 have been measured via exact enumeration
of the $2^N$ spin configurations for small even sizes $10 \leq N \leq 24$,
with a statistics similar to Eq. \ref{ns1copy}.
We have used five small values of the amplitude
$\delta=0.01,0.02,0.03,0.04,0.05$ in order to extract the chaos exponent from
the expansion (Eq \ref{corredefzeta})
\begin{eqnarray}
C_{\delta}(N) \opsimeq_{\delta \to 0} 1-a_{disorder} (\delta N^{\zeta_{J}})^2+o(\delta^2)
\label{corredefzetaJ}
\end{eqnarray}
where $a_{disorder} $ is a numerical constant.
As an example, we show on Fig. \ref{figdisorder} our data for $\sigma=0.75$.

In the non-extensive region $0 \leq \sigma<1/2$,
 our numerical results are compatible
with the value given by Eqs \ref{thetanonext} and \ref{zetaJnonext}
\begin{eqnarray}
 \zeta_J(0 \leq \sigma <1/2)=\frac{1}{2}-\theta(0 \leq \sigma <1/2) \simeq \frac{1}{6} \simeq 0.17
\label{zetaJnonextnum}
\end{eqnarray}

In the extensive region $\sigma>1/2$,
our numerical measures
\begin{eqnarray}
\zeta_J(\sigma=0.62) && \simeq 0.26
\nonumber \\
\zeta_J(\sigma=0.75) && \simeq 0.3
\nonumber \\
\zeta_J(\sigma=0.87) && \simeq 0.33
\nonumber \\
\zeta_J(\sigma=1) && \simeq 0.36
\nonumber \\
\zeta_J(\sigma=1.25) && \simeq 0.44
\label{zetaJnum}
\end{eqnarray}
 yield
the following estimations
 for the surface dimension $d_s(\sigma)= 2 (\theta(\sigma)+\zeta_J(\sigma))$ 
of extensive droplets
(Eq \ref{zetaJext})
\begin{eqnarray}
d_s(\sigma=0.62) && \simeq 1
\nonumber \\
d_s(\sigma=0.75) && \simeq 0.94
\nonumber \\
d_s(\sigma=0.87) && \simeq 0.82
\nonumber \\
d_s(\sigma=1) && \simeq 0.72
\nonumber \\
d_s(\sigma=1.25) && \simeq 0.4
\label{dsJnum}
\end{eqnarray}

\subsection{ Finite disorder perturbation } 

\begin{figure}[htbp]
\includegraphics[height=6cm]{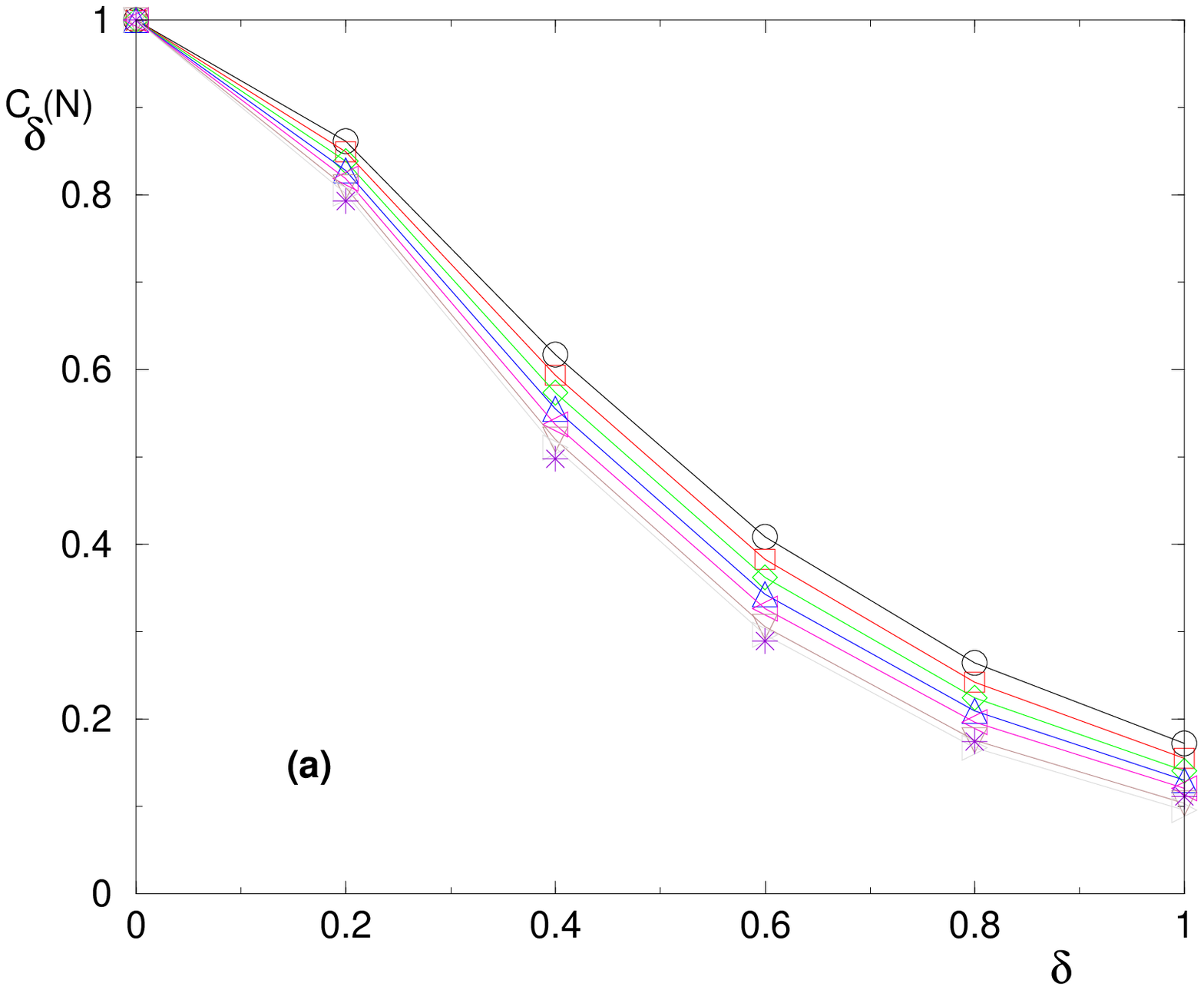}
\hspace{1cm}
 \includegraphics[height=6cm]{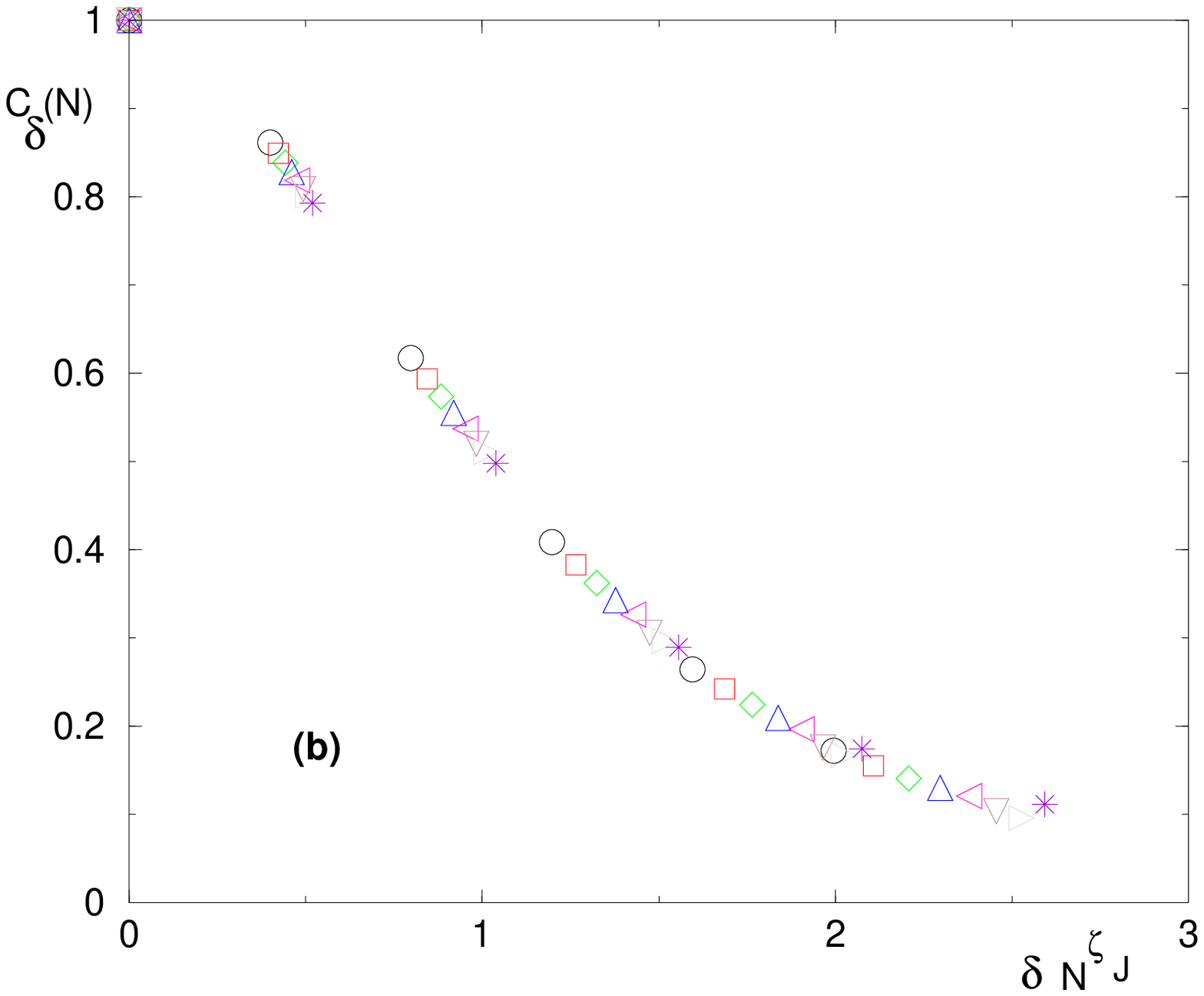}
\caption{ Disorder perturbation of finite amplitude $\delta$ for $\sigma=0.75$ :
(a) Results for the correlation $C_{\delta}(N)$ as a function of the amplitude
$\delta=0.2,0.4,0.6,0.8,1.$ of the perturbation (Eq. \ref{defjijnewdelta} )
for various sizes $10 \leq N \leq 24$.
(b) Same data as a function of the rescaled variable $\delta N^{\zeta_J}$
with $\zeta_J(\sigma=0.75)  \simeq 0.3 $. }
\label{figdisorderFINI}
\end{figure}

We have also study numerically disorder perturbation (Eq. \ref{defjijnewdelta})
 with a finite amplitude $\delta$.
As shown on Fig. \ref{figdisorderFINI} for $\sigma=0.75$, we find that the chaos exponent 
extracted from the expansion of Eq. \ref{corredefzetaJ} for small amplitude $\delta$,
allows to rescale also the results for finite $\delta$.

\section{Temperature chaos exponent $\zeta_T(\sigma)$ }

\label{sec_temp}

\subsection{ Scaling prediction of the droplet theory}

\begin{figure}[htbp]
\includegraphics[height=6cm]{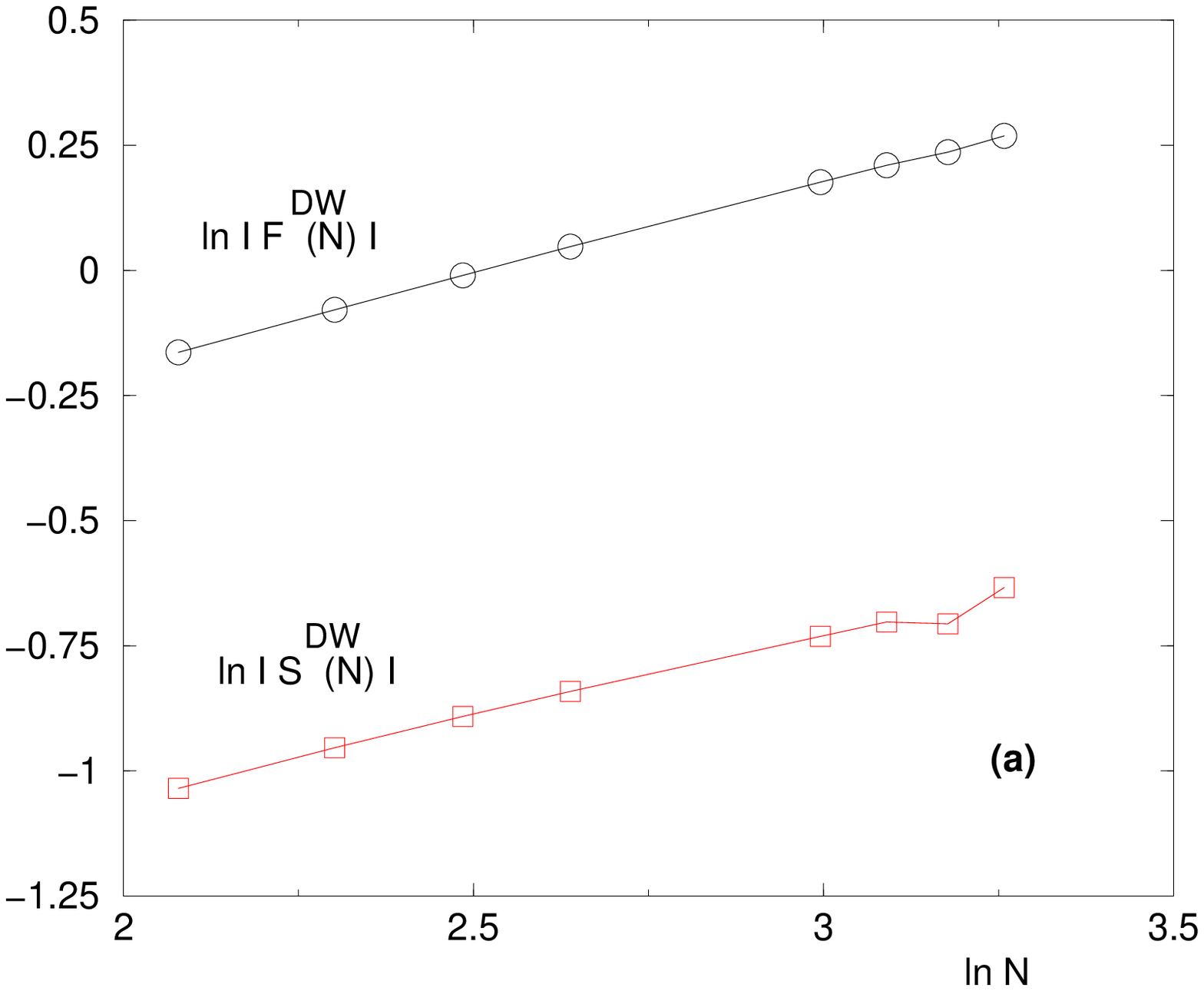}
\hspace{1cm}
 \includegraphics[height=6cm]{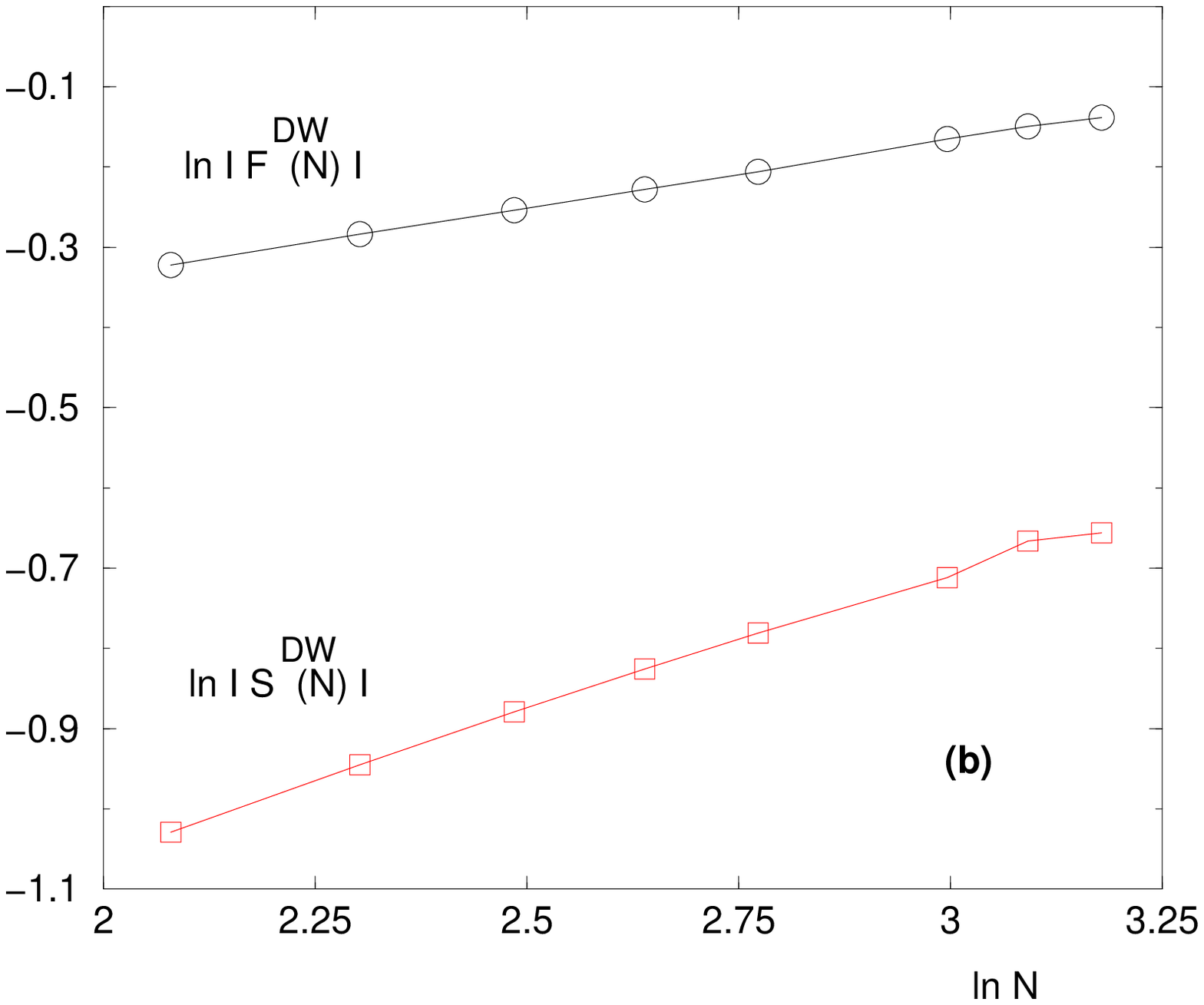}
\caption{ Comparison between the scaling of the Domain-Wall free-energy $F^{DW}(N)$ (Eq. \ref{dwfree}) and the  Domain-Wall entropy $S^{DW}(N)$ (Eq. \ref{dwentropy}) at temperature $T=0.05$ in log-log plots :
(a) for $\sigma=0.25$ in the non-extensive region, the two slopes coincide
 $\theta(\sigma=0.25) \simeq 1/3 \simeq \theta_S(\sigma=0.25)$.
(b) for $\sigma=0.75$ in the extensive region,  the droplet exponent $\theta(\sigma=0.75) \simeq 0.17$ for the free-energy is smaller than the entropy exponent $\theta_S(\sigma=0.75)
\simeq 0.33 $.  }
\label{figentropy}
\end{figure}

Within the droplet scaling theory \cite{bray_moore,fisher_huse},
the chaos exponent associated to a temperature perturbation $\delta T$
for {\it short-range } spin-glasses
can be predicted via the following Imry-Ma argument : 
the perturbation actually involves the same scaling as Eq. \ref{JperSR},
as a consequence of the scaling the entropy of extensive droplets 
as $ L^{\frac{d_s}{2}}$ (coming from some Central Limit Theorem for independent local contributions along the interface)
\begin{eqnarray}
 \Delta^{SR}_T(N) 
\propto (\delta T) L^{\frac{d_s}{2}} = (\delta T) N^{\frac{d_s}{2d}}
\label{TperSR}
\end{eqnarray}
The comparison with the renormalized coupling $J^R(N) \sim N^{\theta} u$
 of Eq. \ref{jrtheta} yields that the appropriate scaling parameter is  
$(\delta T) N^{\zeta_T}$ with the temperature chaos exponent
\begin{eqnarray}
 \zeta^{SR}_T=\frac{d_s}{2 d}-\theta
\label{zetaTSR}
\end{eqnarray}
that coincides with the disorder chaos exponent of Eq. \ref{zetaJSR}.

For the one-dimensional long-range model, the argument about independent local contributions along the interface leading to Eq. \ref{TperSR} cannot be used anymore, and 
we have thus studied numerically
the scaling of the entropy of droplets via the difference of entropy between Periodic and Antiperiodic Boundary conditions
\begin{eqnarray}
S^{DW}(N) \equiv  S^{(P)}(N)-S^{(AP)}(N) \sim  N^{\theta_S} v
\label{dwentropy}
\end{eqnarray}
(where $v$ is an $O(1)$ random variable of zero mean)
that defines the entropy exponent $\theta_S$.
It should be compared with the droplet exponent $\theta$
that governs the free-energy difference of Eq. \ref{jrtheta}
\begin{eqnarray}
F^{DW}(N) \equiv  F^{(P)}(N)-F^{(AP)}(N) \sim  N^{\theta} u
\label{dwfree}
\end{eqnarray}
As examples, we shown on Fig. \ref{figentropy} our results for $\sigma=0.25$ 
and $\sigma=0.75$. Our conclusions are the following :

(i) we find that the entropy exponent $\theta_S(\sigma) $ takes the simple value
\begin{eqnarray}
\theta_S(\sigma) \simeq \frac{1}{3}
\label{thetaS}
\end{eqnarray}
for all $\sigma$. This is actually consistent
 with the same constant value found recently for
the dynamical barrier exponent $\psi(\sigma) \simeq \frac{1}{3}$ \cite{us_dynamic}
(see \cite{us_conjecturepsi} for the conjecture
 on the relation between $\theta_S$ and $\psi$).

(ii) in the non-extensive region $0 \leq \sigma<1/2$ (see Fig. \ref{figentropy} (a)), 
the entropy exponent of Eq. \ref{thetaS} coincides with the droplet exponent
\begin{eqnarray}
\theta_S(\sigma<1/2) \simeq \frac{1}{3}=\theta(\sigma<1/2) 
\label{thetaSnonext}
\end{eqnarray}
so that the corresponding temperature
chaos exponent actually vanishes
\begin{eqnarray}
 \zeta_T(\sigma<1/2)=0
\label{zetaTLRnonext}
\end{eqnarray}

(iii) in the extensive region $\sigma>1/2$ where the droplet exponent is smaller
 $\theta(\sigma)<1/3$
\begin{eqnarray}
\theta_S(\sigma>1/2) \simeq \frac{1}{3}>\theta(\sigma<1/2) 
\label{thetaSext}
\end{eqnarray}
 this means that there exists an entropy-energy cancellation mechanism as
in short-ranged models \cite{bray_moore,fisher_huse}, and that the corresponding temperature
chaos exponent is positive
\begin{eqnarray}
 \zeta_T(\sigma>1/2)=\frac{1}{3}-\theta(\sigma>1/2) >0
\label{zetaTLR}
\end{eqnarray}

\subsection{ Numerical results for the long-range Ising spin-glass }

\begin{figure}[htbp]
\includegraphics[height=6cm]{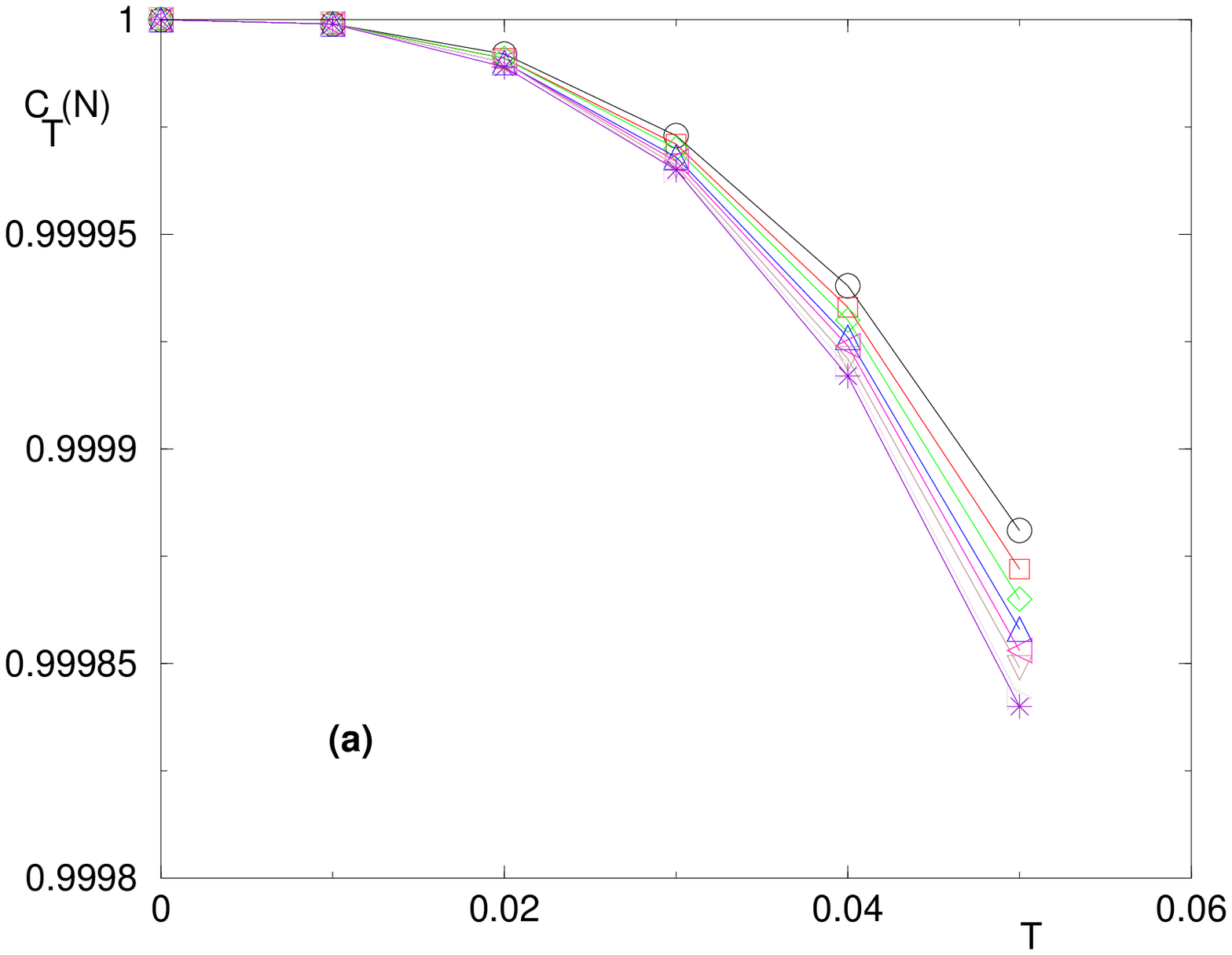}
\hspace{1cm}
 \includegraphics[height=6cm]{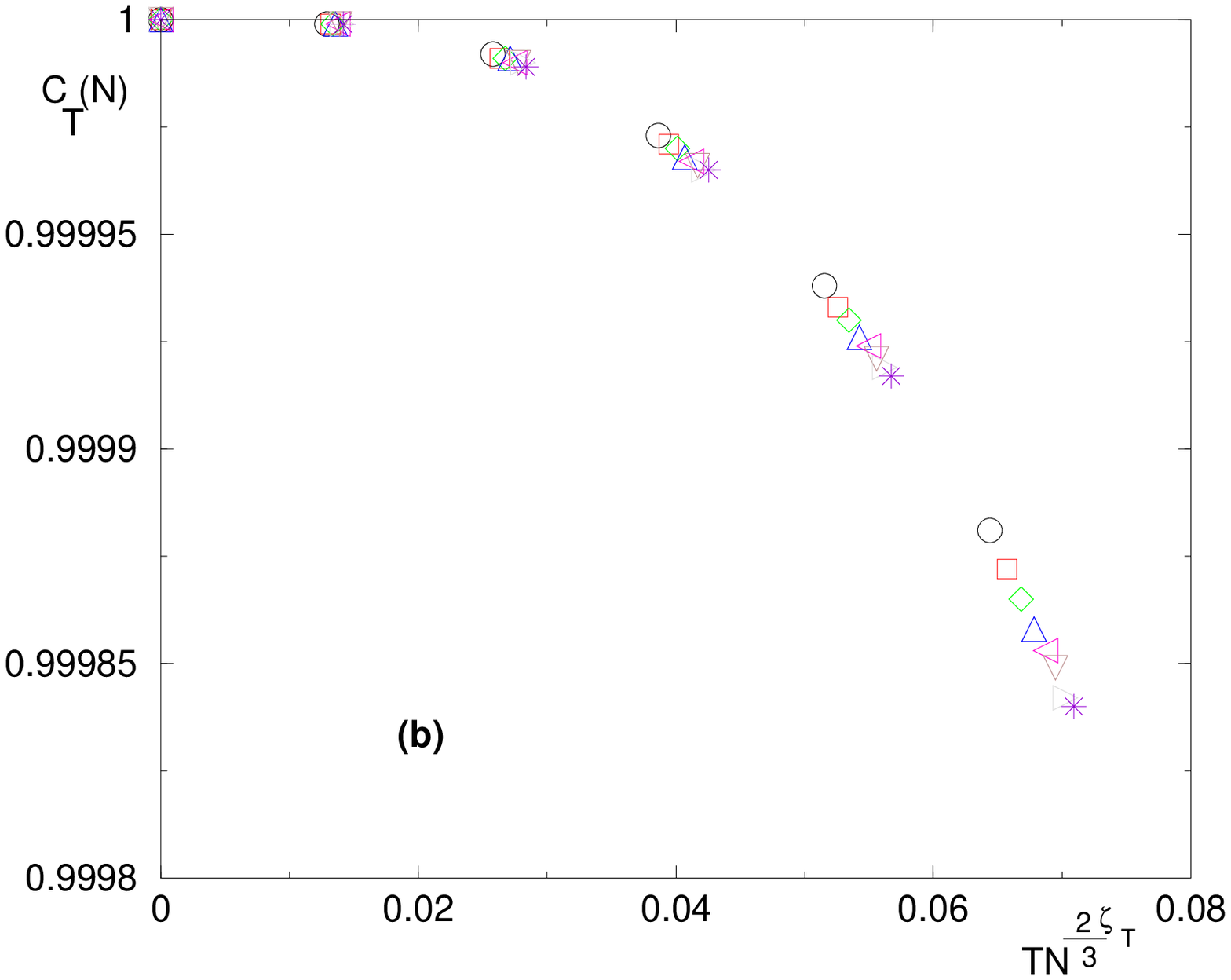}
\caption{ Measure of the temperature chaos exponent $\zeta_T$ for $\sigma=0.75$ :
(a) Results for the correlation $C_{T}(N)$ as a function of the temperature
$T=0.01,0.02,0.03,0.04,0.05$ 
for various sizes $10 \leq N \leq 24$.
(b) Same data as a function of the rescaled variable $T N^{\frac{2}{3} \zeta_T}$
 (see Eq. \ref{corredefzetaT})
with $\zeta_T(\sigma=0.75)  \simeq 0.17 $. }
\label{figtemp}
\end{figure}

We have measured the following correlation (Eq. \ref{corre}) to study temperature perturbations with respect to zero-temperature
\begin{eqnarray}
C_{T}(N) \equiv \frac{ \overline{\left[ F^{(P)}_{T=0}(N)-F^{(AP)}_{T=0} (N)\right]\left[ F^{(P)}_{T}(N)-F^{(AP)}_{T} (N)\right] }}
{\sqrt{\overline{\left[ F^{(P)}_{T=0}(N)-F^{(AP)}_{T=0} (N)\right]^2}}
\sqrt{\overline{\left[ F^{(P)}_{T}(N)-F^{(AP)}_{T} (N)\right]^2}}}
\label{correTzeroT}
\end{eqnarray}
 via exact enumeration
of the $2^N$ spin configurations for small even sizes $10 \leq N \leq 24$,
with a statistics similar to Eq. \ref{ns1copy}.
We have used five small values of the temperature
$T=0.01,0.02,0.03,0.04,0.05$ in order to extract the temperature chaos exponent from
the expansion 
\begin{eqnarray}
C_{T}(N) \opsimeq_{T \to 0} 1- a_{temperature} T (T N^{\zeta_{T}})^2+o(T^3)
\label{corredefzetaT}
\end{eqnarray}
where $a_{temperature} $ is a numerical constant.
Note the additional prefactor of $T$ with respect to the standard quadratic expansion of Eq. \ref{corredefzeta} that can be explained from the behavior of the entropy near zero-temperature \cite{aspelmeierBM}.
As an example, we show on Fig. \ref{figtemp} our data for $\sigma=0.75$.

In the non-extensive regime $0 \leq \sigma <1/2$, we find 
that the temperature chaos exponent vanishes
\begin{eqnarray}
\zeta_T(0 \leq \sigma <1/2) && \simeq 0
\label{zetaTnumnonext}
\end{eqnarray}
in agreement  with Eq. \ref{zetaTLRnonext}.

In the extensive region $\sigma>1/2$,
our numerical measures
\begin{eqnarray}
\zeta_T(\sigma=0.62) && \simeq 0.09
\nonumber \\
\zeta_T(\sigma=0.75) && \simeq 0.17
\nonumber \\
\zeta_T(\sigma=0.87) && \simeq 0.26
\nonumber \\
\zeta_T(\sigma=1) && \simeq 0.36
\nonumber \\
\zeta_T(\sigma=1.25) && \simeq 0.58
\label{zetaTnum}
\end{eqnarray}
are in agreement with Eq. \ref{zetaTLR}.

\section{ Instability of the ground state with respect to perturbations }

\label{sec_overlap}

In this section, we describe our numerical results concerning 
the chaoticity parameter of Eq. \ref{chaoticity} at zero temperature $T=0$
to characterize the instability of the ground-state $S_i^{GS}$ 
with respect to a perturbation $\delta$ via the overlap (Eq. \ref{qdelta})
\begin{eqnarray}
 q^{(T=0)}_{(0,\delta)}(N) \equiv \frac{1}{N} \left\vert \sum_{i=1}^N S_i^{GS(0)} S_i^{GS(\delta)} \right\vert
\label{qdeltazeroT}
\end{eqnarray}
Since there is no thermal fluctuations at zero temperature,
the denominator of Eq. \ref{chaoticity} is unity, so that
 the chaoticity parameter of Eq. \ref{chaoticity} reduces to the disorder-average of Eq. \ref{qdeltazeroT}
\begin{eqnarray}
 r^{(T=0)}_{\delta}(N) \equiv \overline{ q^{(T=0)}_{(0,\delta)}(N) } 
\label{chaoticityzerot}
\end{eqnarray}

\subsection{ Magnetic perturbation at zero temperature }

\begin{figure}[htbp]
\includegraphics[height=6cm]{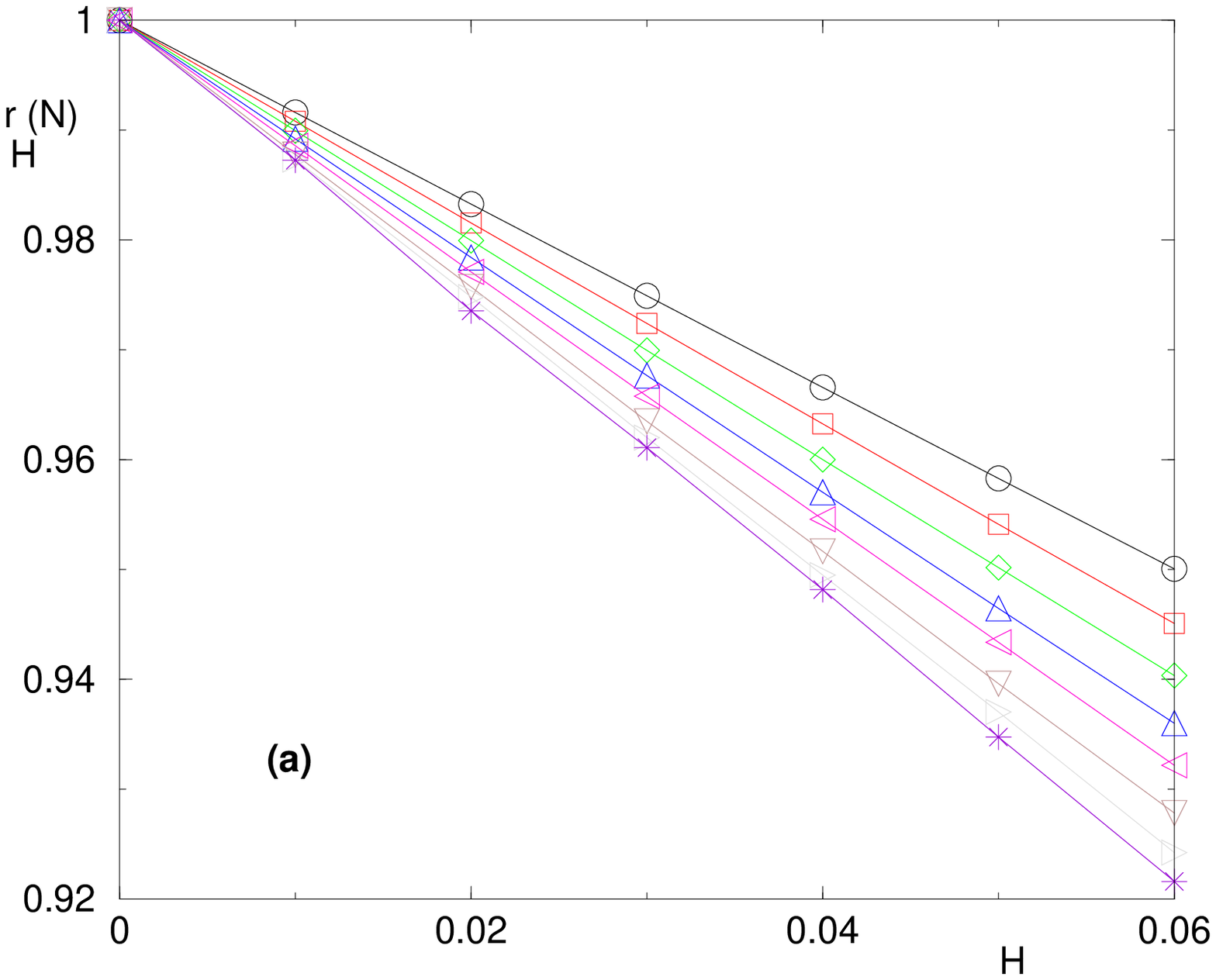}
\hspace{1cm}
 \includegraphics[height=6cm]{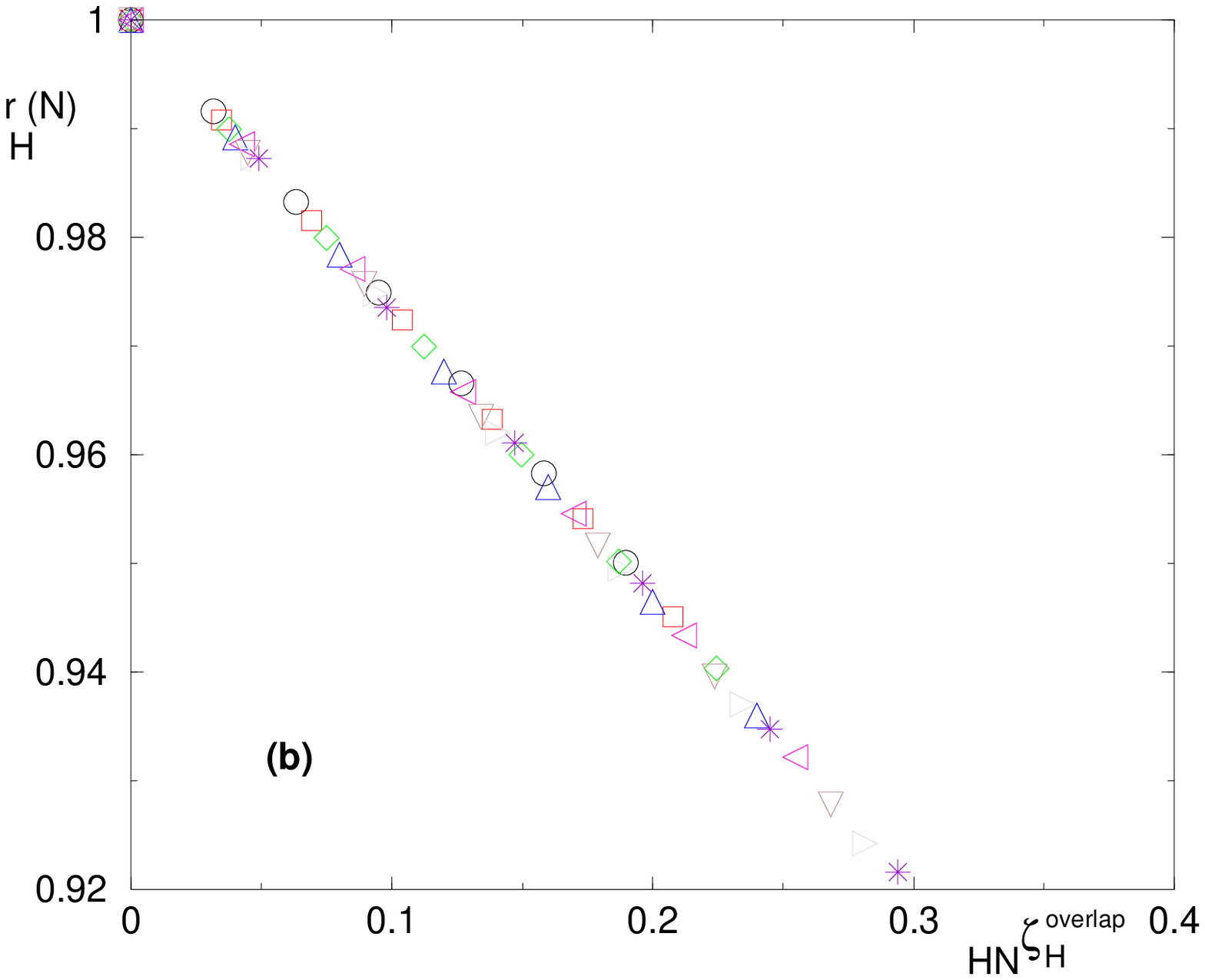}
\caption{ Measure of the chaos exponent $\zeta^{overlap}_H$ for $\sigma=0.75$ :
(a) Results for the chaoticity parameter $r^{(T=0)}_{H}(N)$ 
as a function of the external magnetic field
$H=0.01,0.02,0.03,0.04,0.05,0.06$ 
for various sizes $10 \leq N \leq 24$.
(b) Same data as a function of the rescaled variable $H N^{\zeta_H^{overlap}}$
with $\zeta^{overlap}_H(\sigma=0.75)  \simeq 0.5 $. }
\label{figoverlapmagnetic}
\end{figure}

We have measured 
the chaoticity parameter $r^{(T=0)}_{H}(N)$ of Eq. \ref{chaoticityzerot}.
The ground state configurations of the spins have been obtained
via exact enumeration
of the $2^N$ spin configurations for small even sizes $10 \leq N \leq 24$.
 The statistics over samples is similar to Eq. \ref{ns1copy}.
We have used six small values of the magnetic field
$H=0.01,0.02,0.03,0.04,0.05,0.06$ in order to extract the chaos exponent from
the expansion of Eq. \ref{corredefzetaover}
\begin{eqnarray}
r_{H}(N) \opsimeq_{H \to 0} 1- b_{magnetic} H N^{\zeta^{overlap}_{H}}+o(H)
\label{corredefzetaoverH}
\end{eqnarray}
where $b_{magnetic} $ is a numerical constant.

As an example, we show on Fig. \ref{figoverlapmagnetic} our data for $\sigma=0.75$
leading to 
\begin{eqnarray}
\zeta^{overlap}_H(\sigma=0.75) && \simeq 0.5
\label{zetaoverlapHnum1}
\end{eqnarray}
in agreement with \cite{Katz} (see Fig. 4 of \cite{Katz}).
For $\sigma=0$ corresponding to the mean-field SK model, we also find the same value
\begin{eqnarray}
\zeta^{overlap}_H(\sigma=0) && \simeq 0.5
\label{zetaoverlapHSK}
\end{eqnarray}
in agreement with \cite{parisi,franz,Katz} 
(although the other value $\zeta^{overlap}_H(\sigma=0) \simeq 3/8$ can be found in
\cite{kondor,ritort,franz,barbara}).

\subsection{ Disorder perturbation at zero temperature }

\begin{figure}[htbp]
\includegraphics[height=6cm]{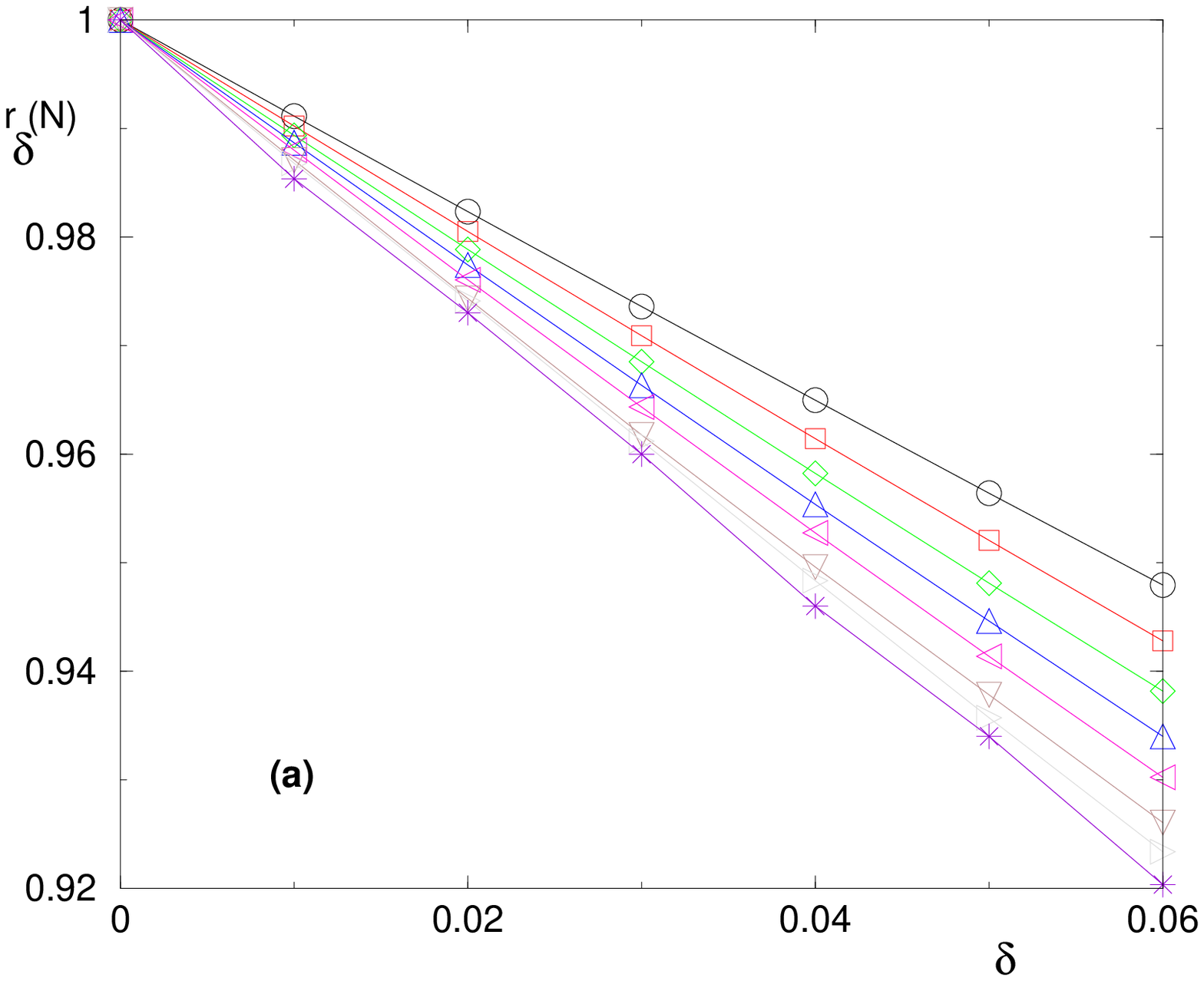}
\hspace{1cm}
 \includegraphics[height=6cm]{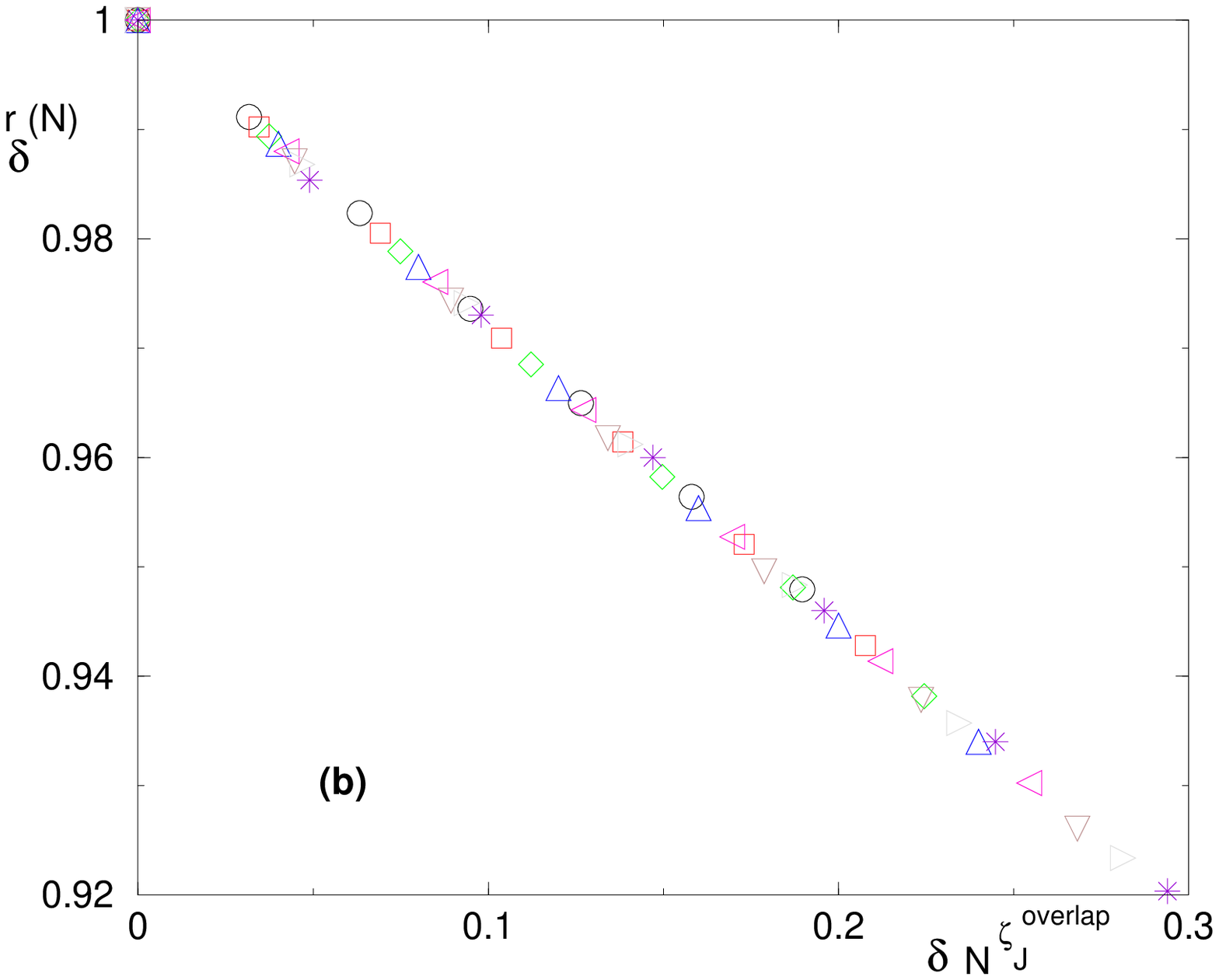}
\caption{ Measure of the chaos exponent $\zeta^{overlap}_{J}$ for $\sigma=0.75$ :
(a) Results for the chaoticity parameter $r^{(T=0)}_{\delta}(N)$ 
as a function of the amplitude
$\delta=0.01,0.02,0.03,0.04,0.05,0.06$ of the disorder perturbation (Eq. \ref{defjijnewdelta})
for various sizes $10 \leq N \leq 24$.
(b) Same data as a function of the rescaled variable $\delta N^{\zeta_\delta^{overlap}}$
with $\zeta^{overlap}_J(\sigma=0.75)  \simeq 0.5 $. }
\label{figoverlapdisorder}
\end{figure}

For the disorder perturbation of Eq. \ref{defjijnewdelta},
we have measured 
the chaoticity parameter $r^{(T=0)}_{\delta}(N)$ of Eq. \ref{chaoticityzerot}.
The ground state configurations of the spins have been obtained
via exact enumeration
of the $2^N$ spin configurations for small even sizes $10 \leq N \leq 24$.
 The statistics over samples is similar to Eq. \ref{ns1copy}.
We have used six small values of the perturbation amplitude
$\delta=0.01,0.02,0.03,0.04,0.05,0.06$ in order to extract the chaos exponent from
the expansion of Eq. \ref{corredefzetaover}
\begin{eqnarray}
r_{\delta}(N) \opsimeq_{\delta \to 0} 1- b_{disorder}  \delta N^{\zeta^{overlap}_{J}}+o(\delta)
\label{corredefzetaoverJ}
\end{eqnarray}
where $b_{disorder} $ is a numerical constant.

As an example, we show on Fig. \ref{figoverlapdisorder} our data for $\sigma=0.75$
leading to 
\begin{eqnarray}
\zeta^{overlap}_J(\sigma=0.75) && \simeq 0.5
\label{zetaoverlapJnum1}
\end{eqnarray}

For $\sigma=0$ corresponding to the mean-field SK model, we also find the same value
\begin{eqnarray}
\zeta^{overlap}_J(\sigma=0) && \simeq 0.5
\label{zetaoverlapJSK}
\end{eqnarray}
in agreement with \cite{ritort2,krzakalaJPB,aspelmeier}.

\subsection{ Explanation in terms of the avalanche triggered by the lowest local field  }

For the mean-field SK model corresponding to $\sigma=0$, the results of Eqs \ref{zetaoverlapHSK} and \ref{zetaoverlapJSK} simply reflects the scaling of the lowest local field $h_{min}(N)
\propto N^{-1/2}$ (see Eq. \ref{hminsk} and explanations in Appendix \ref{app_hlocmin}), since the flipping of the spin corresponding to this lowest local field is known to be able to trigger an extensive avalanche \cite{pazmandi,krzakalaJPB,ledoussal, andresen}.

Our numerical results of Eq. \ref{zetaoverlapHnum1} and \ref{zetaoverlapJnum1} for $\sigma=0.75$ also coincide with the scaling of the lowest local field $h_{min}(N) \propto N^{-1/2}$
(see Eq. \ref{hminsigma} and Figure \ref{figlocalfield0.75} in Appendix \ref{app_hlocmin}).
Our conclusion is thus that for $\sigma=0.75$ also, the flipping of the spin corresponding to the lowest local field is able to trigger an extensive avalanche. 

Note that this is very different from the nearest-neighbor model defined on hypercubic lattices : the flipping of the lowest local field $h_{min}(N) \propto N^{-1}=L^{-d}$ (See Eq. \ref{hmind} and explanations in Appendix \ref{app_hlocmin}) is not able to trigger an extensive avalanche. And the overlap chaos exponents which have been measured in finite $d$ 
with respect to the linear size $L$ are of order 
$ \zeta^{overlap}_J (d=2) \simeq 1$ \cite{rieger,krzakalaJPB}
and $ \zeta^{overlap}_J (d=3) \simeq 1.1$ \cite{krzakalaJPB,katz_krzakala,fernandez}
and are thus much smaller than the value $d$ which would correspond to the scaling of the lowest local field $h_{min}=L^{-d}$.

\section{ Conclusion }

\label{sec_conclusion}

For the long-range one-dimensional Ising spin-glass with random couplings decaying as $J(r) \propto r^{-\sigma}$, we have studied numerically the chaos properties as a function of $\sigma$ for various types of perturbation near the zero-temperature fixed point.

We have first studied the instability of the renormalization flow of the effective coupling defined as the difference between the free-energies corresponding to Periodic and Antiperiodic boundary conditions $J^R(N) \equiv  F^{(P)}(N)-F^{(AP)}(N)$ :

(a) for magnetic perturbations, we have found that the magnetic chaos exponents 
satisfies the standard droplet formula (Eq. \ref{zetaH}) 
involving the droplet exponent $\theta(\sigma)$

\begin{eqnarray}
 \zeta_H(\sigma)=\frac{1}{2}-\theta(\sigma)
\label{zetaHsigma}
\end{eqnarray}

(b) for disorder perturbation, we have measured the disorder chaos exponent
$\zeta_J(\sigma)$, which yields the surface dimension $d_s(\sigma)$ of droplets via the standard droplet formula

\begin{eqnarray}
 \zeta_J(\sigma)=\frac{d_s(\sigma)}{2}-\theta(\sigma)
\label{zetaJsigma}
\end{eqnarray}

(c) for temperature perturbation, we have obtained that the temperature chaos exponent
 $\zeta_T(\sigma)$ satisfies the formula
\begin{eqnarray}
 \zeta_T(\sigma)=\frac{1}{3}-\theta(\sigma)
\label{zetaTsigma}
\end{eqnarray}
where $1/3=\theta_S(\sigma)=\psi(\sigma)$ is the entropic exponent $\theta_S(\sigma) $,
and also the barrier exponent $\psi(\sigma)$ of the dynamics.

 Then we have also studied the instability of the ground state configuration with respect to perturbations, as measured by the spin overlap between the unperturbed and the perturbed ground states. Both for magnetic and disorder perturbations,
we have obtained for all $\sigma$ the exponent
\begin{eqnarray}
\zeta^{overlap}_H(\sigma) =\frac{1}{2}= \zeta^{overlap}_J(\sigma)
\label{zetaoversigma}
\end{eqnarray}
which simply reflects the scaling of the lowest local field (Eq. \ref{hminsigma}) that can trigger an extensive avalanche.

For all these cases, we have discussed the similarities and differences with short range models in finite dimension $d$.

\appendix

\section{ Scaling of the lowest local field $h_{min}(N)$ at zero temperature  }

\label{app_hlocmin}

From the probability distribution $P_{N}(h)$ of the local field
\begin{eqnarray}
h_i \equiv \vert \sum_{j} J_{ij} S_j^{(0)} \vert
\label{hloc}
\end{eqnarray}
seen by spins in the ground-state of a spin-glass model of $N$ sites,
the typical lowest local field $h_{min}(N)$ in a sample
can be estimated from
\begin{eqnarray}
\frac{1}{N}=\int_0^{h_{min}(N)} dh P_{N}(h)
\label{hminphcritere}
\end{eqnarray}

\subsection{ Finite dimension with nearest-neighbor interaction }

For nearest-neighbor spin-glass models defined on hypercubic lattices in dimension $d>1$,
the probability distribution $P_{N}(h)$ has a finite weight at $h=0$ in the thermodynamic limit \cite{boettcher} 
\begin{eqnarray}
 P^{(d)}_{\infty}(h=0) > 0
\label{phfinited}
\end{eqnarray}
so that the lowest local field scales as (Eq. \ref{hminphcritere})
\begin{eqnarray}
h^{(d)}_{min}(N) \simeq \frac{1}{N P_{N=\infty}(h=0) } 
\label{hmind}
\end{eqnarray}

\subsection{ Mean-field SK model }

For the mean-field SK model, the probability distribution $P_{N=\infty}(h)$
vanishes linearly \cite{tap,palmer}
\begin{eqnarray}
P^{(SK)}_{N=\infty}(h) \oppropto_{h \to 0}  h
\label{phsk}
\end{eqnarray}
so that Eq. \ref{hminphcritere} implies the scaling
\begin{eqnarray}
h^{(SK)}_{min}(N)  \oppropto \frac{1}{N^{\frac{1}{2}}}
\label{hminsk}
\end{eqnarray}

\subsection{ Long-range one-dimensional spin-glass model }

\begin{figure}[htbp]
\includegraphics[height=6cm]{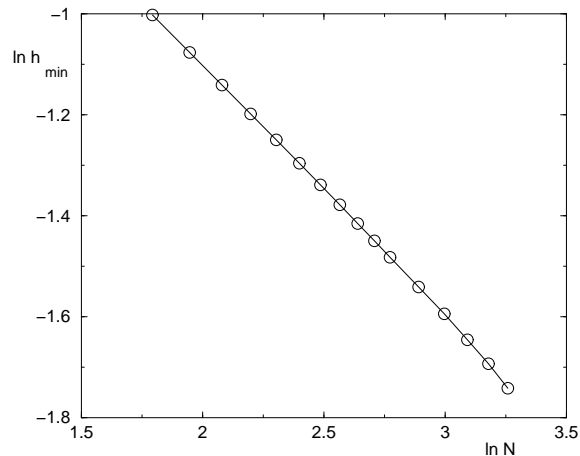}
\caption{ Scaling of the lowest local field $h_{min}(N)$ in the ground state
 for $\sigma=0.75$ as a function of the system size $N$ :
the log-log plot corresponds to the slope $-0.5$, i.e. to the scaling of Eq. \ref{hminsigma}.}
\label{figlocalfield0.75}
\end{figure}

For the long-range one-dimensional spin-glass model,
the probability distribution $P_N(h)$ has been studied numerically in \cite{boettcher}.
Since the behavior of the histogram
$P_N(h) $ near $h=0$ is difficult to extrapolate \cite{boettcher}, we have chosen instead
to study directly the lowest local field $h_{min}$ via exact enumeration of the ground states on the sizes $6 \leq N \leq 26$.
As shown on Fig. \ref{figlocalfield0.75} for the case $\sigma=0.75$,
we find the scaling analogous to Eq. \ref{hminsk} for all values of $\sigma \geq 0$
\begin{eqnarray}
h^{(\sigma)}_{min}(N)  \oppropto \frac{1}{N^{\frac{1}{2}}}
\label{hminsigma}
\end{eqnarray}
Note that for the histogram, this corresponds to the finite-size behavior
\begin{eqnarray}
P^{(\sigma)}_{N}(h=0) \propto \frac{1}{N^{\frac{1}{2}}}
\label{phfiniten}
\end{eqnarray}
 via Eq. \ref{hminphcritere}.

\end{document}